\documentclass[fdp,a4paper,fleqn,amsmath,amssymb]{w-art}

\usepackage{amsbsy}
\usepackage{amsmath}
\usepackage{amsthm}

\usepackage{times,cite,w-thm}

\theoremstyle{plain}

\theoremstyle{definition}

\usepackage[english]{babel}
\usepackage[]{graphicx}

\newcommand{\dm}{n}
\newcommand{\be}{\begin{equation}}
\newcommand{\ee}{\end{equation}}
\newcommand{\ba}{\begin{eqnarray}}
\newcommand{\ea}{\end{eqnarray}} 
\newcommand{\M}{\mathcal{M}}

\newcommand{\A}[1]{A^{\!(#1)}}
\newcommand{\eps}{\varepsilon}
\newcommand{\cv}[1]{{\partial}_{#1}}
\newcommand{\KV}[1]{\xi^{(#1)}}

\begin{document}
\DOIsuffix{theDOIsuffix}

%%    Dates will be filled in by the publisher. The 'reviseddate' and
%%    'dateposted' (Published online) entry may be left empty.
\Receiveddate{XXXX}
\Reviseddate{XXXX}
\Accepteddate{XXXX}
\Dateposted{XXXX}

%% \pretitle{Editor's Choice}

%% We have a short and a long form for the title. The short form
%% (optional argument) goes into the running head.

\title[Hidden Symmetries and Integrability]{Hidden Symmetries and Integrability in Higher Dimensional Rotating Black Hole Spacetimes}

%% Please do not enter footnotes or \inst{}-notes into the optional
%% argument of the author command. The optional argument will go into
%% the header.  If there is only one address the marker \inst{x} may be
%% omitted.

%% Information for the first author.
\author[M. Cariglia]{Marco Cariglia\inst{1,}%
  \footnote{E-mail:~\textsf{marco@iceb.ufop.br}.}}
\address[\inst{1}]{Universidade Federal de Ouro Preto, ICEB, Departamento de F\'isica.
  Campus Morro do Cruzeiro, Morro do Cruzeiro, 35400-000 - Ouro Preto, MG - Brasil}
%%
%%    Information for the second  author
\author[P. Krtou\v{s}]{Pavel Krtou\v{s}\inst{2,}\footnote{E-mail:~\textsf{Pavel.Krtous@utf.mff.cuni.cz}.}}
\address[\inst{2}]{Institute of Theoretical Physics, Faculty of Mathematics and Physics, Charles University \\
V~Hole\v{s}ovi\v{c}k\'ach 2, Prague, Czech Republic}
%%
%%    Information for the third author
\author[D. Kubiz\v n\'ak]{David Kubiz\v n\'ak\inst{3,}\footnote{E-mail:~\textsf{dkubiznak@perimeterinstitute.ca}.}}
\address[\inst{2}]{Perimeter Institute, 31 Caroline St. N. Waterloo Ontario, N2L 2Y5, Canada}

\begin{abstract}
This is a short pedagogical introduction to the subject of Killing-St\"{a}ckel and Killing--Yano tensors and their role in the integrability of various types of equations that are of physical interest in curved space-time, the main application being higher dimensional rotating black holes with cosmological constant. 
\end{abstract}

%% maketitle must follow the abstract.
\maketitle                   % Produces the title.

\section{Introduction}
Much of recent years' research in theoretical physics has been devoted to understanding the link between quantum and gravitational physics. Different themes occur in this line of work, to mention some: string theory and the AdS/CFT correspondence, supergravity and supersymmetric theories, black holes and the study of their microscopic degrees of freedom and their stability.

In this brief report we are going to sketch the role of Killing--St\"{a}ckel and Killing--Yano (KY) tensors in the study of (quantum) physics in curved space-time, and provide some useful references. On one hand these objects have a direct relationship with the idea of symmetry of motion in a curved spacetime: the seminal work \cite{Carter1968} of Carter in 1968 showed that geodesic trajectories in the Kerr geometry possess an unexpected constant of motion that is not related to isometries. It is related to a symmetry of dynamics that acts upon the full phase space but has no direct action on configuration space alone: a so called 'hidden symmetry'. Hidden symmetries are in correspondence with Killing--St\"{a}ckel tensors.

On the other hand, Floyd \cite{Floyd1973} and Penrose \cite{Penrose1973} showed that the Kerr Killing--St\"{a}ckel tensor admits a tensorial square root, a KY tensor, and Carter and McLenaghan \cite{CarterMcLenaghan1979} showed it can be used to build a differential operator that commutes with the Dirac operator in this metric, thus proving how special tensors can be used to study quantum mechanical problems for fields of non-zero spin. There is also an interesting connection with supersymmetric theories, as an appropriate semi-classical limit of the Dirac equation is the theory of the point particle with $\mathcal{N}=1$ worldline supersymmetry. In this theory rank two KY tensors are associated to generators of additional supersymmetries  \cite{GibbonsRietdijkVanHolten}.

In recent years interest in higher dimensions coming from string theory, braneworld models, study of black holes' hair, motivated the study of hidden symmetries in dimension greater than four. The Kerr-NUT-(A)dS black hole spacetime \cite{ChenLiuPope2006} which is a generalization of the Kerr metric to higher dimensions, has horizon with spherical topology and is parametrized by spin, NUT charges and a cosmological constant. Therefore it lends itself to an analysis first as a black hole in its own merit, then in terms of a possible Kerr/CFT correspondence \cite{StromingerEtAl2011}, and also in terms of the AdS/CFT duality. It is remarkable that such geometry possesses many of the important properties of the original Kerr metric: it has a KY tensor \cite{DavidFrolov2006} and the Hamilton--Jacobi, Klein--Gordon and Dirac equations are separable \cite{DavidPavelFrolov2007, OotaYasui:2008}. It is also possible to construct conserved quantities in this metric for the theory of the spinning particle that generalize the original one found by Carter to the case of non-zero spin \cite{DavidMarcoSpinningParticle}. 

The study of KY tensors can be generalized by looking for these special tensors in different types of metrics, see for example \cite{Papadopoulos2008} for a connection between KY tensors and G-structures, or by generalizing the KY equation to metrics with fluxes, see 
\cite{Hari:2009, DavidEtAlFlux2010,DavidEtAlFlux2010_2,Papadopoulos2011} for applications. 

Thus it is manifest how KY tensors link diverse topics such as black hole physics, special geometry, integrability, supersymmetry and supergravity, CFT type dualities; see also recent extended review \cite{Yasui:2011}. 

%%%%%%%%%%%%%%%%%%%%%%%%%%%%%%

\section{Phase space and hidden symmetries\label{sec:hiddensymmetries}}
Let a manifold $\M$ be a configuration space of dimension $\dm$, its co-tangent bundle $T^* (\M)$ be the phase space with coordinates $y^a=\{x^\mu, p_\nu\}$,  and $\omega =\omega_{ab} dy^a \wedge dy^b=dx^\mu\wedge dp_\mu$ be the natural symplectic form, with inverse components $\omega^{ab}$. Given a scalar Hamiltonian function $H:T^* (\M) \mapsto \mathcal{R}$, the equations of motion are written as 
\be 
\frac{d y^a}{d \tau}  = \omega^{ab} \partial_b H := X_H^a \, . 
\ee
The vector $X_H$ is the \textit{symplectic gradient} of the function $H$. 
%Also given $f,g$ functions  valued in $T^* (\M)$ we can define Poisson brackets 
%\be 
%\left\{ f , g \right\} = \omega^{ab} \partial_a f \partial_b g = X_g (f) = - X_f(g) \, . 
%\ee
Symplectic gradients are in one-to-one correspondence with (local) canonical transformations, which are  (local) continuous transformations of $T^* (\M)$ generated by a vector field $X$ such that the symplectic form is left invariant, $\mathcal{L}_X \omega = 0$. So in particular the form of Hamilton's equations is invariant and volume in phase space is invariant. 

Now consider a Killing vector $K^\mu$: the phase-space quantity $C_K = K^\mu p_\mu$ is conserved as it can be checked that $\dot{C}_K = \left\{C_K, H \right\} =0$ due to the Killing equation. Its corresponding symplectic gradient is 
\be 
X_{C_K} = K^\mu \frac{\partial}{\partial x^\mu} - \frac{\partial K^\lambda}{\partial x^\mu} p_\lambda \frac{\partial}{\partial p_\mu} \, . 
\ee
The pushforward of $X_{C_K}$ under the canonical projection $\pi: T^* (\M) \rightarrow \M$ is given by $\pi_* (X_{C_K}) = K^\mu \frac{\partial}{\partial x^\mu} = K$. 
The geometrical interpretation is that the integral curves of $X_{C_K}$ in phase space have a projection on configuration space, and this is given by the integral curves of $K$ which are the isometries.

Next, a Killing--St\"{a}ckel  tensor is totally symmetric tensor $K^{\mu_1 \dots \mu_p} = K^{(\mu_1 \dots \mu_p)}$ such that $\nabla^{( \lambda} K^{\mu_1 \dots \mu_p)} = 0$, which implies that $C_K = K^{\mu_1 \dots \mu_p} p_{\mu_1} \dots p_{\mu_p}$ is conserved. The symplectic gradient of $C_K$ is 
\be 
X_{C_K} = p K^{\mu_1 \dots \mu_{p-1}\nu} p_{\mu_1}  \dots p_{\mu_{p-1}} \frac{\partial}{\partial x^\nu} - \frac{\partial K^{\mu_1 \dots \mu_p}}{\partial x^\nu} p_{\mu_1} \dots p_{\mu_p} \frac{\partial}{\partial p_\nu} \, . 
\ee
It still generates a symmetry of phase space but this time the push forward to configuration space vanishes: $\pi_* (X_{C_K}) = 0$. So there is no visible action of the flux of $X_{C_K}$ in configuration space. It is said that $K$ generates a {\em hidden symmetry} of the dynamics. New non-trivial Killing--St\"{a}ckel tensors of order $\ge 3$ have recently been built in \cite{GibbonsEtAl:2011, GibbonsEtAl_2:2011}.

%%%%%%%%%%%%%%%%%%%%%%%%%%%%%%

\section{Killing--Yano tensors}
The Killing equation can be generalized alternatively using a totally symmetric tensor $f_{\mu_1 \dots \mu_p} = f_{[\mu_1 \dots \mu_p]}$. The differential equation satisfied by $f$ is 
\be \label{eq:KY}
\nabla_\lambda f_{\mu_1 \dots \mu_p} + \nabla_{\mu_1} f_{\lambda \dots \mu_p} = 0 \, .  
\ee
When $p=1$ this reduces to the Killing equation. Such a tensor is called a {\em Killing--Yano tensor}, and it was first introduced by Yano \cite{Yano:1952} from a purely mathematical point of view. The tensor 
\be 
K_{\mu\nu} = f_{\mu \lambda_1 \dots \lambda_{p-1}} f_\nu{}^{\lambda_1 \dots \lambda_{p-1}} \, ,
\ee
can be seen to be Killing--St\"{a}ckel, using Eq. \eqref{eq:KY}. In order to be able to generate a conserved quantity using $f$ we cannot contract $f$ with factors of the momenta $p_\mu$, we need appropriate antisymmetric variables: this can be done in the theory of the spinning particle and for the Dirac equation. For the latter it is possible to show that a KY tensor always allows to construct an operator that commutes with the Dirac operator, with no quantum anomaly \cite{Cariglia:2004}. 
KY tensors can also be used to construct conserved gravitational charges \cite{Kastor:2004}.

There exists a natural way to generalize the KY equation to an equation that is invariant under Hodge duality. A tensor satisfying this latter equation is called a {\em conformal Killing--Yano tensor}, and its square is a conformal Killing--St\"{a}ckel tensor.  Since a KY tensor has zero divergence, then its Hodge dual is a  closed form, called {\em closed conformal Killing--Yano tensor}. Such tensors satisfy a very useful property: they form an algebra under the wedge product. In particular, closed conformal Killing--Yano tensors of rank 2 that are non-degenerate are called \textit{Principal conformal Killing--Yano (PCKY) tensors}. They are crucial for the integrability of various systems in four and higher dimensional black hole spacetimes.

%%%%%%%%%%%%%%%%%%%%%%%%%%%%%% 

\section{Kerr-NUT-(A)dS black holes} 
% Kerr-NUT-(A)dS black holes  admit a PCKY tensor .
While a classification of Lorentzian metrics with a PCKY tensor is not available, the analogue problem in Riemannian signature has been solved \cite{HouriOotaYasui2007,KrtousFrolovKubiznak2008}. The most general {\em canonical metric} admitting a PCKY tensor in $\dm = 2N + \eps$ dimensions, $\eps=0,1$, is given by 
\be  \label{metric}
ds^2
  = \sum_{\mu=1}^N\biggl[ \frac{d x_{\mu}^{\;\,2}}{Q_\mu}
  +Q_\mu\Bigl(\,\sum_{j=0}^{N-1} \A{j}_{\mu}d\psi_j \Bigr)^{\!2}  \biggr]  + \eps S \Bigl(\,\sum_{j=0}^N \A{j}d\psi_j \Bigr)^{\!2}.
\ee
Here, coordinates $x_\mu\, (\mu=1,\dots,N)$ stand for the (Wick rotated) radial coordinate and longitudinal angles, and Killing
coordinates $\psi_k\; (k=0,\dots,N-1 +\eps)$ denote time and azimuthal angles associated with Killing vectors
${\KV{k}} =\cv{\psi_k}$. We have further defined the functions 
\ba
Q_\mu&=&\frac{X_\mu}{U_\mu}\,,\quad U_{\mu}=\prod\limits_{\nu\ne\mu} (x_{\nu}^2-x_{\mu}^2)  \;,\quad S = \frac{-c}{\A{N}} \, ,\label{eq:UandS_def}\\
\A{k}_{\mu}&=&\hspace{-5mm}\!\!
    \sum\limits_{\substack{\nu_1,\dots,\nu_k\\\nu_1<\dots<\nu_k,\;\nu_i\ne\mu}}\!\!\!\!\!\!\!\!\!\!
    x^2_{\nu_1}\cdots\, x^2_{\nu_k}\;,\ \
\A{j} = \hspace{-5mm} \sum\limits_{\substack{\nu_1,\dots,\nu_k\\\nu_1<\dots<\nu_k}}\!\!\!\!\!\!
    x^2_{\nu_1}\cdots\, x^2_{\nu_k}\; .\label{eq:A_def}\quad
\ea
The quantities ${X_\mu}$ are functions of a single variable ${x_\mu}$, and $c$ is an arbitrary constant.
The vacuum (with a cosmological constant) black hole geometry is recovered by setting
\begin{equation}\label{BHXs}
  X_\mu = \sum_{k=\eps}^{N}\, c_{k}\, x_\mu^{2k} - 2 b_\mu\, x_\mu^{1-\eps} + \frac{\eps c}{x_\mu^2} \; .
\end{equation}
This choice of $X_\mu$ describes the most general known Kerr-NUT-(A)dS spacetimes in all dimensions \cite{ChenLiuPope2006}. The constant $c_N$ is proportional to the cosmological constant and the remaining constants are related to angular momenta,
mass and NUT parameters. 

The PCKY tensor reads \cite{DavidFrolov2006}
\be
h=db\,,\quad b=\frac{1}{2}\sum_{j=0}^{N-1}A^{(j+1)}d\psi_j\,.
\ee 
$2j$-forms $h^{(j)}$, which are the $j$-th wedge power of the PCKY tensor $h$, $h^{(j)} = h\wedge \dots \wedge h$, form the tower of associated closed conformal Killing--Yano tensors, and can be `squared' to rank 2 Killing--St\"{a}ckel tensors.

\section{Results on integrability for the canonical spacetimes}
The canonical metric has a tower of $N+\eps$ Killing vectors ${\KV{k}}$ and $N$ rank 2 Killing--St\"{a}ckel tensors $K^{(i)ab}$. It can be shown that these satisfy the geometric compatibility conditions to form a \textit{separability structure} as discussed in \cite{BenentiFrancaviglia1979,KalninsMiller1984}, and as a consequence both the Hamilton--Jacobi and Klein--Gordon equations are separable. Separability for both equations was obtained with a direct calculation in \cite{DavidPavelFrolov2007}, then in the case of Hamilton--Jacobi it was associated to the separability structure in \cite{HouriOotaYasui2008}, and last for the Klein--Gordon equation it was shown to be associated to a set of commuting differential operators of first and second order associated to, respectively, Killing vectors and Killing tensors, that are simultaneously diagonalizable \cite{PavelSergyeyev2008}: 
\be 
\mathcal{L}^{(k)} = -i \xi^{(k) a} \nabla_a \, , \qquad \mathcal{K}^{(i)} = - \nabla_a \left[ K^{(i) ab} \nabla_b \right] \, . 
\ee
The Nambu-Goto equations for a stationary string were integrated in \cite{DavidFrolov2008}. 

When looking at the Dirac equation, a theory of necessary and sufficient conditions for separation of variables is not available. Separation of variables was obtained by explicit calculation in \cite{OotaYasui:2008}. In \cite{CarigliaKrtousKubiznak2011} it was shown that the canonical metric admits a set of $\dm$ mutually commuting operators defined on the Clifford bundle, one being the Dirac operator $D$. The other ones divide into $N+\eps$ operators constructed from the Killing vectors $\xi^{(k)}$ and $N-1$ operators constructed form $2j$-forms $h^{(j)}$, according to 
\begin{align}
&\begin{aligned}
  K_k&=  \xi^{(k) a} \nabla_a +\frac{1}{4}\gamma^{a_1 a_2} \nabla_{[a_1} \xi^{(k)})_{a_2]} \, , 
\end{aligned}\\
&\begin{aligned}
  M_j&= \frac1{(2j)!}\,\Bigl[\gamma^{a b_1\dots b_{2j}}(h^{(j)})_{b_1\dots b_{2j}}\nabla_a -\frac{2j(n{-}2j)}{2(n{-}2j{+}1)}\gamma^{b_1\dots b_{2j{-}1}}(\delta h^{(j)})_{b_1\dots b_{2j{-}1}}\Bigr]\, . 
\end{aligned}
\end{align}
These operators admit a common eigenfunction which displays separation of variables \cite{CarigliaKrtousKubiznak2011_2}, and coincides with that found in \cite{OotaYasui:2008}. 

Recently in \cite{DavidMarcoSpinningParticle} conserved quantities of order two in the momenta have been constructed for the theory of the spinning particle: these generalize the Killing--St\"{a}ckel charges of Sec. \ref{sec:hiddensymmetries} to the case of non-zero spin.

\begin{acknowledgement}
M.C.  would like to thank the organizers of the "XVII European Workshop on String Theory", held in September 2011 in Padova, Italy, and 
acknowledges partial financial support from Funda\c c\~ao de Amparo \`a Pesquisa de Minas Gerais - FAPEMIG.
\end{acknowledgement}

% Use this code if you wish to generate your bibliography with BibTeX;
% please replace first the string "demo" below with the name(s) of
% the BibTeX data base(s) you want to use.
% The resulting bibliography-output (the contents of the .bbl file)
% must be pasted into this file before submission.
% 
% \bibliographystyle{pss}
% \bibliography{demo}
% 
% Replace the following example bibliography with your references
% before submission:

\end{document}